\documentclass[twocolumn,showpacs,preprintnumbers,amsmath,amssymb]{revtex4}
 
\usepackage{graphicx}
\usepackage{dcolumn}
\usepackage{bm}
 
\def\pb{$\overline{p}\;$}
\def\nb{$\overline{n}\;$}

\begin{document}
 
\preprint{Report LPSC-03-06}

\input epsf
\title{Secondary proton flux induced by cosmic ray interactions with the atmosphere}

\author{B. Baret, L. Derome, C.Y. Huang\footnote{Present address: MPIK, Saupfercheckweg 1, 
D-69117 Heidelberg, Germany}, and M. Bu\'enerd\footnote{Corresponding author: buenerd@in2p3.fr}} 
\affiliation{Laboratoire de Physique Subatomique et de Cosmologie, IN2P3/CNRS, 53 Av. des Martyrs, 
38026 Grenoble-cedex, France}

\date{\today}

\begin{abstract}
The atmospheric secondary proton flux is studied for altitudes extending from sea level 
up to the top of atmosphere by means of a 3-dimensional Monte-Carlo simulation procedure 
successfully used previously to account for flux measurements of protons, light nuclei, 
and electrons-positrons below the geomagnetic cutoff (satellite data), and of muons and 
antiprotons (balloon data). The calculated flux are compared with the experimental 
measurements from sea level uo to high float ballon altitudes. The agreement between 
data and simulation results are very good at all altitudes, including the lowest ones,
where the calculations become extremely sensitive to the proton production cross section. 
The results are discussed in this context. The calculations are extended to the study of 
quasi trapped particles above the atmosphere to about 5 Earth radii, for prospective 
purpose.
\end{abstract}

\pacs{94.30.Hn,95.85.Ry,96.40.-z,13.85.-t}%
\maketitle 
\section{Introduction}
The study of the particle flux in the earth neighbourhood has regained some interest 
recently with the emergence of a new generation of embarked experiments both at balloon 
and satellite altitudes. In this context, new measurements of the proton flux have been 
performed \cite{FR99,FU01}, providing a broad set of accurate data which could be compared 
to the latest generation of calculations. This is a quite compelling test which concerns 
the atmospheric flux of all secondary particles as well, since the latter are driven by 
$p$ induced cross sections along the atmospheric cascade.
The interaction dynamics of the incident Cosmic Ray (CR) flux with the earth atmosphere 
and the earth magnetic field, for secondary particle production, is a complex 
process. So far it has been investigated only by means of a theoretical approach based on
a diffusion equation \cite{PA96}. A new detailed study of this process through the body of
recently measured data should significantly improve the current status of the knowledge 
in this matter, and it should validate the calculations based on this, or similar, simulation 
procedure  for the evaluation of all the atmospheric secondary particle flux. It is also 
likely to improve our knowledge on the dynamics of the population of the radiation belts 
as well.

Studying the secondary proton flux in the atmosphere in this context is therefore of 
particular interest since it is highly sensitive to all the components of the simulation 
process, in particular to the secondary proton production cross section as discussed below. 
A successful account of this flux through the range of atmospheric altitudes is then likely 
to cast robust grounds to this approach in general, and to further validate the computation 
techniques used. Studying this flux at higher altitudes in the Earth environment on the 
basis establised previously should also be a useful investigation both from the point of 
view of the particle dynamics and for the future satellite experiments for which this 
background must be known. 

The present work is a further step of a research program whose previous results on the flux 
of secondary atmospheric particles at satellite altitude, in particular on the interpretation 
of the AMS 98 measurements of protons, leptons, and light ions, below the geomagnetic cutoff 
(GC) \cite{PAP1,PAP2,PAP3},
and in the atmosphere (muons and neutrinos) \cite{LI03}, have been reported recently.
The investigation of the proton flux reported in \cite{PAP1} is extended here
to the atmospheric altitudes and to the high altitude region. The article reports on the 
calculated atmospheric $p$ flux from sea level to balloon altitude, on the comparison to the 
experimental data, and on the predictions of the flux at high altitudes from top of 
atmosphere (TOA) up to 3 10$^4$~km. The main features of the calculations are described in 
section \ref{SIMU}. 
The production cross sections used in the event generator are briefly discussed in 
section~\ref{CROSEC}. 
The simulation results are discussed in section~\ref{RESP}, while the transport equation 
approach is described and the results are shown in section~\ref{EQDIF}. The proton flux at 
high altitudes is reported in \ref{HIGHALT}. The work is concluded in section~\ref{CONC}. 

The paper parallels a similar study on the antiproton flux in the earth environment 
\cite{HU03}, referred to as I in the following. The two papers are presented in this order
for historical reasons.
\section{Simulation conditions}\label{SIMU}
The flux of secondary atmospheric protons has been investigated using the same simulation 
approach used in \cite{PAP1,PAP2,PAP3,LI03} and for the antiproton (see I) flux in the 
atmosphere. 
The same computer code has been used here for the charged particle propagation in the 
terrestrial environment including the atmosphere, as in the previous studies.
Incident CR proton and helium particles were generated and propagated inside the earth 
magnetic field, interacting with atmospheric nuclei according to their total reaction 
cross section and producing secondary nucleons $p$, $n$, light nuclei, leptons 
($\mu,e^\pm,\nu$) from meson ($\pi, K$) decay, and antinucleons \pb, \nb, with 
cross sections and multiplicities as discussed below. 
Each secondary particle produced in a given collision is propagated in the same conditions 
as incident CRs in the previous step, resulting in a more or less extended reaction cascade 
developing through the atmosphere, which included up to about ten generations of secondaries 
for the protons of the simulation sample \cite{PAP1}. 

The reaction products are counted each time they are crossing, upwards or downwards, the 
virtual detection spheres. The locations of the latter were chosen between sea level and about 
36~km for ground and balloon experiments (BESS, CAPRICE), at 380~km for the AMS satellite 
experiment, and beyond up to about 3 10$^4$~km for the rest of the study. All charged 
particles undergo energy loss by ionisation in the atmospheric medium. 
Each event is propagated until the particle disappears by nuclear collision, stopping in 
the atmosphere by energy loss, or escaping to the outer space beyond twice the generation 
altitude (see \cite{PAP1,PAP2,PAP3} and I).

The incident CR proton and helium flux have been measured recently by several experiments 
\cite{BO99,ME00,SA00,PROT,HELI}. In the present work, functional form fits to the AMS 
data \cite{PROT,HELI} have been used in the calculations to generate the corresponding flux.
For other periods of the solar cycle than those of the measurements, the incident cosmic flux 
were corrected for the different solar modulation effects using a simple force law 
approximation.
The $A>4$ components of the CR flux were not taken into account in the present calculations
(see ref~\cite{LI03} for details).
\section{Cross sections}\label{CROSEC}

The inclusive $p+A\rightarrow p+X$ and $^4He+A\rightarrow p+X$ proton production cross 
sections and the proton and $^4$He total reaction cross sections on nuclei used here are 
described in I. For incident protons however, the description of the production cross section
below 7.5~GeV \cite{BA85} used in \cite{PAP1}, has been constrained by the 4.2~GeV measurements 
from ref~\cite{AG93}. This decreased the cross section over this range of energy by a few tens
of percents. The details will be reported later.

The neutron production cross sections were taken the same as for protons. This is expected to be
a good approximation for the incident and final state energies considered here. Charge-exchange
reaction channels have been neglected in account of their small cross sections. The nucleon 
production from $He$ fragmentation were not included.

\section{Results}\label{RESP}
\begin{figure}[!htb]          
\begin{center}
\includegraphics[width=8cm]{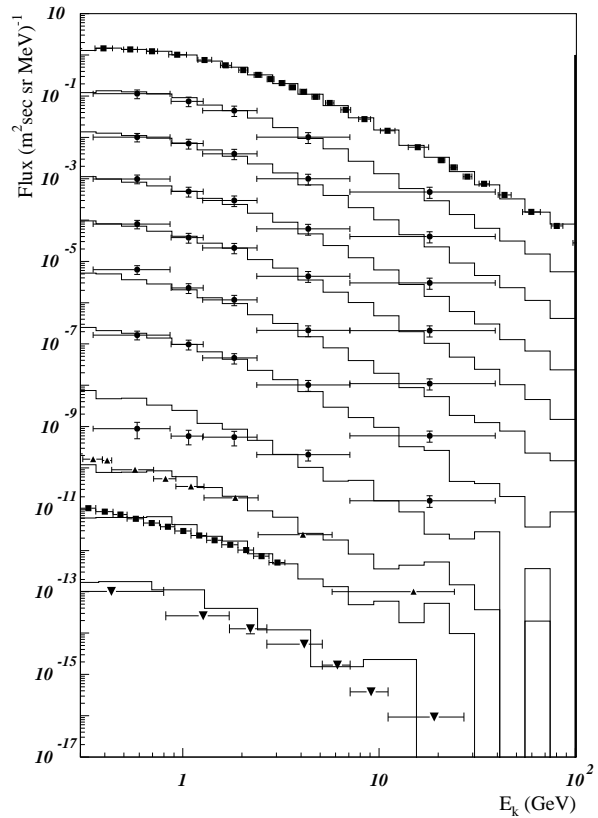} 
\vspace{-1cm} 
\caption{\it\small Simulated proton flux in the atmosphere (histograms) compared to experimental 
results (symbols) between TOA and sea level. From top to bottom: CAPRICE TOA \cite{BO99} (full 
squares), CAPRICE 94 \cite{FR99} (7 altitudes: 29.9, 22.1, 17.5, 16, 12.6, 9.9, 5.75, in km), 
Kocharian et al., \cite{KO56} (3.2~km, full triangles), BESS \cite{FU01} (2.77km, full squares), 
Diggory et al., \cite{DI74} (sea level, inverted full triangles). See also \cite{BR64} for other 
data at sea level. 
\label{PROTS}}
\end{center}
\end{figure}
\begin{figure*}[!htb]          
\begin{center}
\includegraphics[width=12cm]{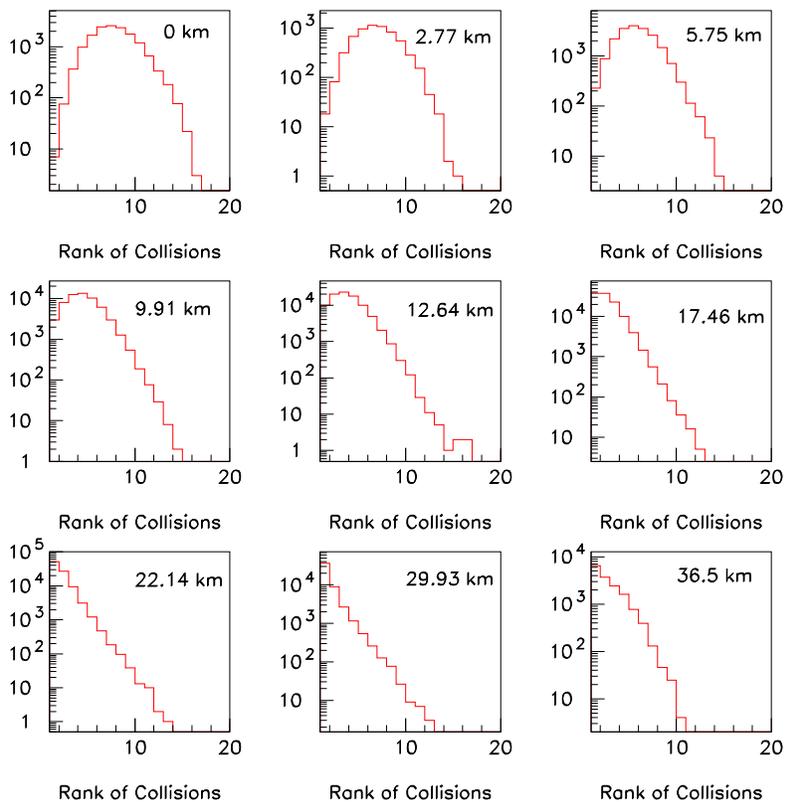} 
\caption{\it\small Rank distributions of the detected protons obtained from the simulation, 
for a set of altitudes between sea level and balloon high float altitude, showing the evolution
of the mean production rank in the collision sequence, of the detected protons.
\label{RANKS}}
\end{center}
\end{figure*}

The proton flux in the atmosphere have been measured recently by the CAPRICE experiment 
between 5 and 29.9~km \cite{FR99} and at TOA \cite{BO99}, and by BESS at lower altitude 
(2.77~km) \cite{FU01}, while previous measurements were available from \cite{KO56} at 3.2~km, 
and from \cite{DI74} at sea level.
This collection of data points is compared on figure~\ref{PROTS} with the simulation results 
between sea level up to balloon altitudes. The comparison is remarkably good through the whole 
range of altitudes. Note that the calculations become increasingly sensitive to the proton 
production cross-section when going from high to low altitudes since all secondary protons 
result on the average from a sequence of $n$ collisions $N+A\rightarrow p+X$, $N$ standing 
for nucleon and $n$ being the average collision rank, involving (approximately) the same 
inclusive nucleon production cross section for each collision. 
Let the latter be noted as $\frac{d\sigma_N(E_i,\theta,E)}{d\theta dE}$ for a given 
collision with $E_i$, $E$, $\theta$, being the nucleon incident energy, proton final energy,
and production angle, respectively. Although the collisions involve different values of the 
kinematic variables, the final proton flux for a given altitude depends qualitatively on some 
appropriate average of the cross section above to the $n$th power, i.e.,: 
$<\!\frac{d\sigma_N(E_i,\theta,E)}{d\theta dE}\!>^{n}$, with $n$ increasing with the decreasing 
altitude. Therefore the lower the altitude, the more sensitive the flux is expected to be to 
the inclusive proton production cross section.

Fig~\ref{RANKS} shows a sample of rank distributions from sea level to ballon altitude 
obtaines from the simulation sample. The mean rank $n$ appears to drop from around 7 at sea 
level down to about 1.5 at 36.5~km. The flux is thus sensitive to approximately the 7th power of
the cross section at sea level, while it depends only to the 3/2 power of this cross section,
approximately, at ballon altitude. A discrepancy of a factor of 2 between calculation results 
and data at sea level can thus be considered as a good result since it would point to a 
discrepancy on the cross section of less than 10 percent. 

These calculations thus provide an extremely sensitive test of the correctness of the cross 
sections used, and beyond of the overall method, and they assign a high confidence level to 
the all secondary particle flux calculated by this approach.

\section{The transport equation approach}\label{EQDIF}
The proton flux has also been calculated using the same diffusion equation approach as used 
in refs \cite{PF96} (see also \cite{PA96,ST97,ST93} ) in order to compare the two methods. 

\subsection{Numerical integration method}
The equation to be solved can be written as:
\begin{eqnarray}
\lefteqn{ \frac{\partial N_i}{\partial x} +
 \frac{\partial}{\partial E_i}\bigg\langle \frac{\partial E_i}{\partial x}\bigg\rangle 
 N_i(x,E_i) + \frac{N_i(x,E_i)}{\lambda_{int}(E_i)} }\nonumber\\
& {}- \sum_{A}\frac{1}{\langle m_{air}\rangle}\int_{E_{th}}^{\infty}
    \frac{d\sigma}{dE_i}(E_i,E_{A})N_{A}(x,E_{A})dE_{A}=0
\end{eqnarray}
Where $i$ represents the transported particle species, $N_i(x,E_i)$ its energy ($E_i$)
dependent flux after crossing the thickness between $x$ and $x+dx$ (in $g/cm^{2}$)of 
(atmospheric) matter. The second term is the particle energy changing term accounting for 
energy loss by ionization (using the Bethe-Bloch formula).  In the third (absorption) term, 
$\lambda_{int}$ is the particle interaction length derived from the total reaction cross 
section of the considered system. The last term is the source term, accounting for particle 
creation, with $A$ standing for the projectile-target system leading to the production 
of particle $i$. 
$\frac{d\sigma}{dE_i}(E_i,E_A)$ is the production cross section for particle $i$ in the 
system $A$ with incident energy $E_A$. The sum runs over all the allowed $A$ channels. 
The differential cross sections used here are the same (angle integrated) as used in the 
simulation. $\langle m_{air}\rangle$ is the mean nuclear mass of the atmospheric nuclei 
(14.58 amu).

The equation has been solved for the secondary proton flux at various altitudes. The 
technique used proceeds by the method of finite differences with implicit scheme proposed
in ref.~\cite{RECIP}. It consists of solving a system of linear equations: 

{\setlength\arraycolsep{2pt}
\begin{eqnarray}
N_{n-1}^{m}&=&\frac{\Delta x}{\Delta E^{m-1}+\Delta E^{m}} \bigg\langle 
\frac{\partial E_i}{\partial x} \bigg\rangle^{m}(N_{n}^{m+1}-N_{n}^{m-1})+ \nonumber\\ 
& & (1+\frac{\Delta x}{\lambda^{m}})N_{n}^{m}
 -\frac{\Delta x}{\langle m_{air}\rangle}\sum_{m'=1}^{M_{MAX}}\bigg(
\frac{d\sigma^{m}}{dE}(E_{bin}^{m'})+ \nonumber\\
& & \frac{d\sigma^{m}}{dE}(E_{bin}^{m'+1})\bigg)
\frac{\Delta E^{m'}}{2}N_{n}^{m'}
\end{eqnarray}}
%
%
%
Where $m$ and $n$ are the index of the steps in energy and thickness crossed respectively. 

For practical reasons related to the energy range to be covered versus the number of steps
required and the size of the matrix to be inverted, the definition of the derivative used at 
the boundaries had to be modified to obtain stable calculation results \cite{RECIP}. 

\subsection{Results}
\begin{figure}[htb]          
\begin{center}
\includegraphics[width=8cm]{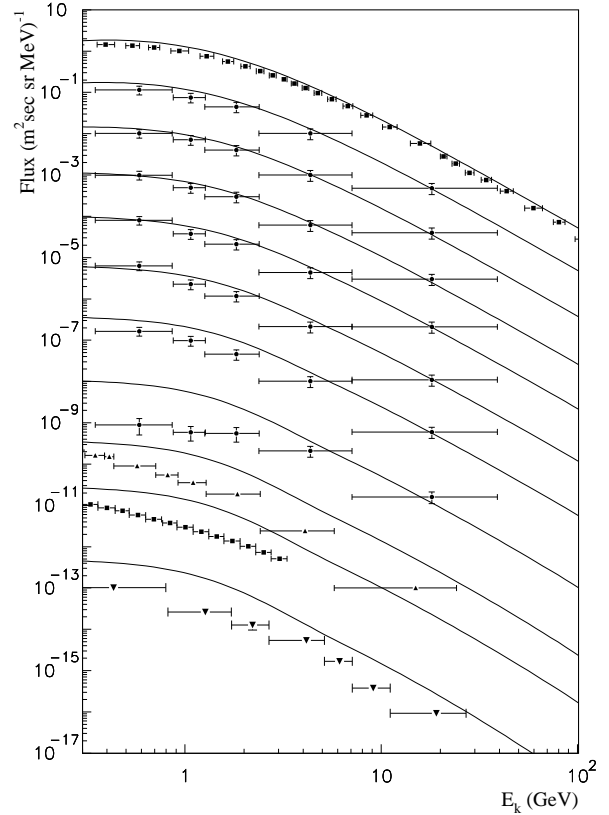} 
\caption{\it\small Results obtained using the diffusion equation approach for the transport of
the proton flux in the atmosphere. The data are the same as in Fig.~\ref{PROTS}
\label{DIFF}}
\end{center}
\end{figure}
Fig.~\ref{DIFF} shows the results obtained in these calculations, compared to the 
same experimental data as in Fig.~\ref{PROTS}, at various altitudes in the atmosphere. The 
agreement is seen to be fair in the high altitude range, down to around 10~km. For the 
5.5~km data, the observed disagreement is similar to that obtained by simulation, the data 
points having a much different energy behaviour than observed for the higher and lower energy
spectra. 

For altitudes lower than 5~km, significant disagreements between data and calculations appear, 
the latter overestimating the data by a factor of about 2 to 4. This disagrement probably 
originates in the one-dimension approximation of this approach. The difference between the 
3-dimensional (3-D) approach of the simulation and the 1-dimension diffusion equation lies in a few
3-D effects not included in the latter approximation : a) The curvature of the particle in the 
magnetic field. b) The angle dependence of the particle production, 1-D approximation using 
the angle integrated cross section. A consequence of the above is that the effective path of 
particles in the atmosphere is longer than assumed in the 1-D approximation. Using the more
realistic values of the path obtained in the simulation, in the diffusion equation brings only
minor improvements to the observed disagreement. Introducing some cuts in the angular range of
integration of the cross section distorts the resulting spectrum and does not improve the 
discrepancy either.

It is interesting to note that the failure of the 1-D approximation reported above at low 
atmospheric altitudes seems not to have a significant effect on the goodness of this 
approximation for the calculation of the atmospheric neutrino flux. This conclusion was 
reached on the basis of the fair agreement obtained between the 3-D and 1-D calculations in 
Ref.\cite{LI03}. This can be understood by the fact that the low altitude neutrino 
production represents less than 10\% of the whole atmospheric production (fraction decreasing 
with the decreasing neutrino energy). 

%
\begin{figure*}[!htb]          
\begin{center}
\includegraphics[width=8cm]{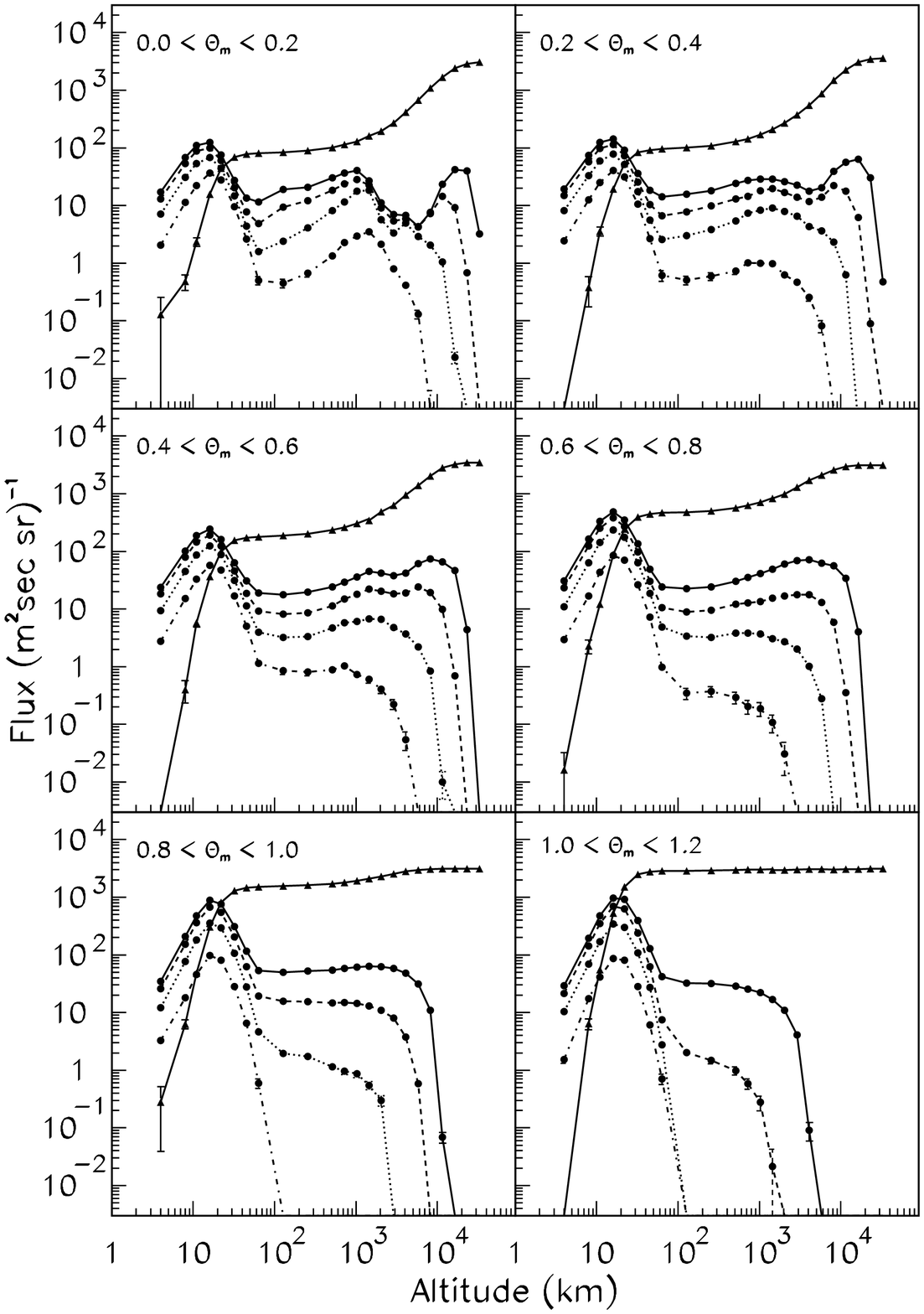} \includegraphics[width=8cm]{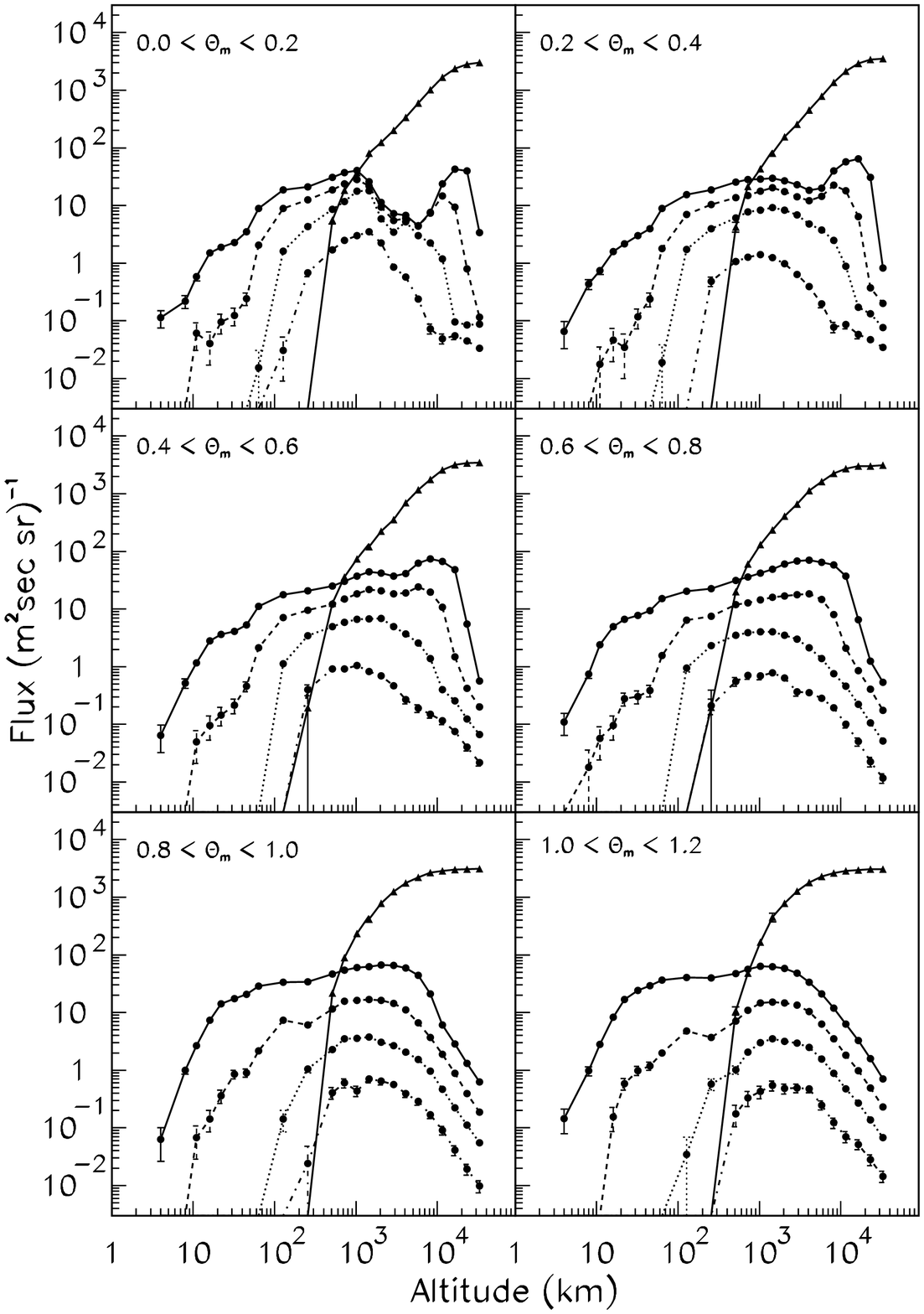} 
\caption{\it\small Simulation results for the altitude dependence of the downgoing 
(left) and upgoing (right) proton flux. Each set of curves of the two panels corresponds 
to a bin of geomagnetic latitude $\theta_m$ as indicated on the figure. The points and 
curves show the energy integrated primary flux of protons (full triangles), and the flux 
for proton with kinetic energies above 0.1~GeV (solid line), 0.3~GeV (dashed line), 1~GeV 
(solid line), and 3~GeV (dash-dotted line), respectively. See text for the discussion. 
\label{PHALT} }
\end{center}
\end{figure*}

\section{Secondary proton flux versus altitude}\label{HIGHALT}

\begin{figure}[!htb]          
\begin{center}
\vspace*{2.0mm} 
\includegraphics[width=8cm]{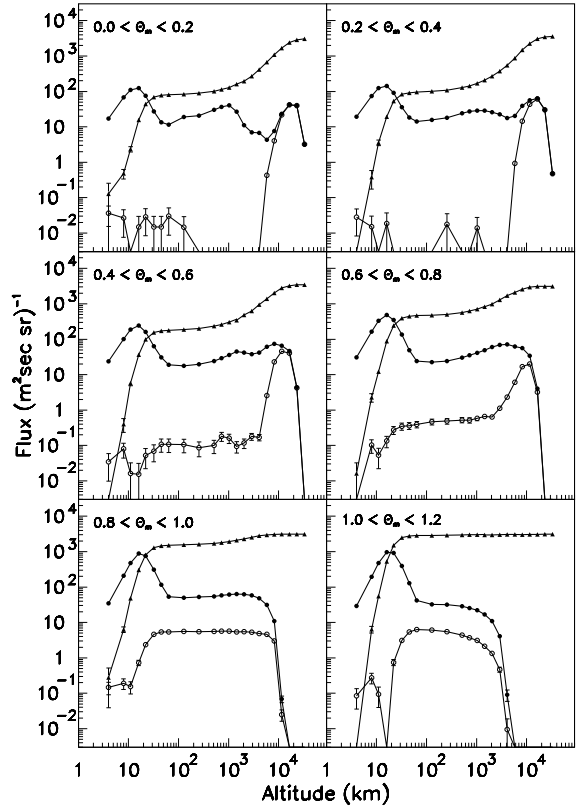} 
\caption{\it\small Same as figure~\ref{PHALT} for the energy integrated downward 
proton flux (full triangle) and the E$>$0.1~GeV (full circles and solid line). The full 
circles symbols and dashed line show the fraction of the latter for which the first 
adiabatic invariant conservation is violated.
\label{ADINV}}
\end{center}
\end{figure}
The same study of the particle flux at high altitudes as performed in I for antiprotons 
has been conducted here for protons over a range of altitudes going from TOA up to 
around 3 10$^4$~km, with the same purpose of, on one hand, understanding better the 
dynamics of the proton population at high altitudes, and on the other hand to provide 
reliable predicions of the atmospheric secondary proton flux for future embarked 
experiments.

Note that the calculations reported in the following include only secondary protons produced
in the atmosphere. They do not include the contribution originating from the atmospheric 
production of neutrons decaying into protons. This contribution is negligible inside the 
atmosphere, but it is a well known major component of the radiation belts for the altitudes 
considered below. This latter issue is investigated in a separate study.

Fig.~\ref{PHALT} left shows the angle ($\theta_z<\pi/4$~rad) and energy (E~$>$~0.1~GeV) 
integrated downgoing primary proton flux from sea level up to 3.2 10$^4$~km, in bins of 
geomagnetic latitudes between equator and polar region.  
At low latitudes, the energy integrated flux is observed to 
drop by more than one order of magnitude between asymptotic distance from Earth and 
about 10$^3$~km  because of the geomagnetic cutoff which is most effective at these 
latitudes. In the polar region where there is no GC, the flux is predicted 
constant over the same range of altitude until it reachs the atmosphere. Inside the 
atmosphere it drops exponentially due to particle absorption.
The secondary flux distributions are shown on the same figure for different energy 
integration thresholds. The main features of these distributions clearly do not depend 
critically on the energy threshold (0.1 to 3~ GeV), although significant differences 
are observed, discussed in the following. The broad peak at low altitude corresponds to 
the atmospheric production 
yield, the low altitude side of the peak being governed by atmospheric absorption, and 
the high altitude side by the atmospheric density. 
The intermediate plateau of the flux distribution corresponds to the population of 
quasi-trapped particles which are accomplishing a few rides between mirror points 
before being absorbed in atmosphere (see discussion in I). The peak observed at high 
altitude around 20-30~10$^3$~km for the low latitude region, corresponds to low energy 
particles for which the first adiabatic invariant is violated (see Fig.~\ref{ADINV}), 
and thus drifting to higher shells without being rapidly absorbed in the atmosphere on 
the normal trajectory between mirror points (see examples in I).
All distributions have a high altitude cutoff, which is both energy and latitude 
dependent, dropping from 3 10$^4$~km for low energy low latitude particles, down to 
10$^2$-10$^3$~km at high energy and high latitude. This latter feature can be 
understood qualitatively by simply considering that: \\
a) Particles produced at low latitudes have a natural momentum limitation set by the 
simple condition that their gyration radius is smaller than the distance of the gyration 
center (mean field line) to the atmosphere. \\
b) Particles produced in the polar region will tend to escape in account of the low 
value of the magnetic field at the poles, and subsequently, the higher the energy, 
the more effective the trend. \\
Quantitatively, this high energy cutoff can be understood in the St\"ormer approach to the 
problem \cite{ST55} which provides the maximum energy allowed for a trapped particle at 
given altitude and latitude. For trapped particle, the predicted dependence of this GC on 
the distance, on the energy, and on the latitude, completely accounts for the features 
observed in Fig.~\ref{PHALT}.

Integrating the zenith angles over a larger range of acceptance produces distributions
significantly wider, extending to lower altitudes, as it could be expected from the above
considerations. 

This is further illustrated on Fig.~\ref{PHALT} right which shows the same distributions for
the upgoing proton flux, complementing the previous figure. The upward primary flux begins 
around 300~km in altitude at the upper boundary of the forbidden region where primary 
trajectories are not allowed by the GC \cite{ST55}. It converges asymptotically with the 
incoming (downward) primary flux at the highest altitude calculated (primary flux isotropy). 
For the upgoing secondaries, a similar plateau and high altitude peak at low latitudes, are 
observed as for the incoming secondary flux. 
These parts of the flux distributions closely overlap with those of the incoming flux at 
low and intermediate latitudes and for the lower energy particles, showing that it is a 
flux of quasi-trapped particles. 
At high latitude the high energy flux which is limited to the inner atmosphere for downgoing 
particles, is to the opposite located at high altitude above 100~km, i.e., approximately 
above TOA, for protons above 3~GeV. These are escape particles which can be produced only 
tangentially to the atmosphere - since forward produced, the backward production cross 
section being vanishingly small - to have a chance of being deflected upwards by the 
magnetic field (with a strong east-west asymmetry effect, see \cite{PAP2}). Consequently, 
the particle trajectories of this flux have a large zenith angle on the average.

Fig.~\ref{ADINV} shows three distributions with the same latitude binning as on the previous 
figures: The angle and energy integrated downgoing proton flux from sea level up to 
3.2 10$^4$~km (full triangles), the energy integrated (E~$>$~0.1~GeV) secondary proton flux 
(full circles, solid line), these two distributions already seen before, and the fraction 
of the secondary flux for which the conservation of the first adiabatic invariant (magnetic 
momentum of the particle \cite{ST55}) is violated by more than one order of magnitude. For the 
low latitudes and high altitudes above about 10$^3$~km, the secondary flux is almost 
exclusively adiabatic invariant violating, while the conservation of the invariant is 
approximately satisfied for quasi-trapped particles over the range of altitudes between 
10$^2$~km and 10$^3$~km.
\section{Summary and conclusion}\label{CONC}
In summary, it has been shown that the simulation approach to the proton flux in the 
atmosphere allows to successfully reproduce the data to a high level of accuracy. This 
result confirms the reliability of the method.
The proton flux has been calculated up to around 10 Earth radii. The results have shown 
that a large component of quasi-trapped particles dominates the flux over the intermediate
range of altitudes (10$^2$-10$^4$~km). These results should serve as a guideline for the
evaluation of the particle background for experiments embarked on satellites. 

A study of trapped protons originating from the neutron flux is in progress, in the same context.
%
%

%
\end{document}